# Image Edge Detection based on Swarm Intelligence using Memristive Networks

Zoha Pajouhi and Kaushik Roy

*Abstract*—**Recent advancements in the development of memristive devices has opened new opportunities for hardware implementation of non-Boolean computing. To this end, the suitability of memristive devices for swarm intelligence algorithms has enabled researchers to solve a maze in hardware. In this paper, we utilize swarm intelligence of memristive networks to perform image edge detection. First, we propose a hardware-friendly algorithm for image edge detection based on ant colony optimization. Second, we implement the image edge detection algorithm using memristive networks. Furthermore, we explain the impact of various parameters of the memristors on the efficacy of the implementation. Our results show 28% improvement in the energy compared to a low power CMOS hardware implementation based on stochastic circuits. Furthermore, our design occupies up to 5x less area.**

## I. INTRODUCTION

Bio-inspired computing has attracted a wide range of interest in the past few years for solving class of problems that are not well suited in von-Neumann architectures [1-6]. Implementation of such biological systems in standard Complementary Metal Oxide Semiconductor (CMOS) devices has turned out be energy inefficient; the inefficiencies stem from both CMOS devices and the computing platform. As an example, let us consider the simulation of cat's brain on IBM's Blue Gene supercomputer. The supercomputer consumes 8 megawatts while the cat's brain consumes only 20 watts [3]. Furthermore, the supercomputer runs two to three orders of magnitude slower than the cat's brain [3].

We believe that proper matching of devices to the algorithms can potentially lead to large improvements in energy consumption. In the quest to achieve comparable power consumption with those of biological counterparts, research has started in earnest to develop newer devices with characteristics similar to biological elements [1-6]. Furthermore, researchers are exploring new computing models to suit bio/neuro-computing systems. Interestingly, the discovery of memristive devices has provided unprecedented similarity between electronic devices and some biological components and has enabled efficient implementation of bio-inspired algorithms [1-2].

Specifically, researchers have demonstrated similarities between memristive networks and *swarm intelligence* algorithms [8-12]. Swarm intelligence is the collaborative behavior of decentralized self-organized agents. These agents work simultaneously and communicate indirectly to find a solution to their problem. One of the most prominent swarm intelligence algorithms is the *ant colony optimization* method [13-16]. It has been shown that the ant colony algorithm is capable of efficiently finding optimal solutions to NP-complete problems such as the traveling salesman problem [13].

Ant colony algorithm mimics the behavior of ants to find food sources. Ants do not possess a sense of sight; however, through efficient, yet simple collaboration, they find the shortest path that leads to food sources. In order to understand the ant colony algorithm, let us consider a simple shortest path problem with two paths as illustrated in Fig. 1 (a). If point A is the ant nest and point B is the food source, there are two different paths to traverse from A to B. In order to find the food source, initially, ants start randomly taking different paths. To start with, roughly half of the ants take path 1 and the other half take path 2. Once they find the food source, they go back home and lay a trail of *pheromones* on their traversal path. The pheromone stays on the path for a certain amount of time and eventually evaporates. In our example, once the ants reach point B, they go back to their home, half of them go through path 1 and half go through path 2; however, since the ones going through path 1 get to their nest sooner, they lay pheromone on the path faster compared to path 2.

Note, ants favor the paths with more pheromone on them over the ones that have less pheromone. Therefore, gradually, the shortest path becomes more alluring to other ants. On the contrary, the pheromones on the longer paths evaporate leaving them less attractive to other ants [13]. Eventually, all the ants take path 1 and the pheromone on path 1 becomes much larger than the pheromone on path 2.

Note that ants do not communicate with each other directly and on a one to one basis; however, they communicate through the pheromone that is laid on the path. This type of communication is called location-based communication. In other words, each path has a memory and remembers the traversal of the ants. A memristive device is a two terminal device that changes its resistance as current passes through it.

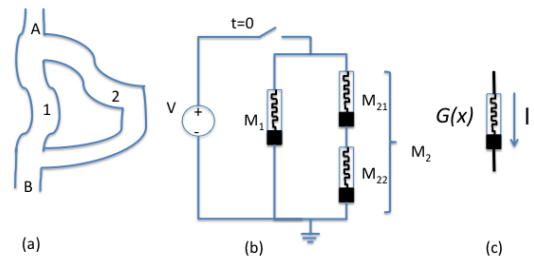

Fig. 1. (a) Points A (nest) and point B (food source) are connected through two paths L1 and L2, such that L2=2L1. (b) Memristive network model of the ant colony model in (a).

For example, let us consider a simple model of a memristive device as illustrated in Fig. 1 (c) [9]:

$$G(x) = G_{on} * x + G_{off} * (1-x), \frac{dx}{dt} = K * I(t), G_{on} > G_{off}, \quad 0 \le x \le 1 \quad (1)$$

Where $G(x)$ is the conductance of the memristive device, $G_{on}$ is the minimum conductance and $G_{off}$ is the maximum conductance. Also, $K$ is the drift factor of the device and $x$ is its internal state. Furthermore, $I(t)$ is the current that passes through the device. If there is no current, the device keeps its current state. However, if a current passes through the device, the internal variable changes based on Eq. 1. For example, if $I(t)$ is a constant value, the internal variable ($x$) increases or decreases linearly based on the direction of the current.

On the other hand, let us consider a memristive network as shown in Fig. 1(b). If we consider the electrons in the circuit similar to ants and the current flow similar to ant traversal in the ant colony algorithm, they may traverse two paths in the circuit: the left path with 1 memristor and the right path with 2 memristors in series.

Furthermore, let us consider that the conductance of all the memristors is $G_{off}$ ($x=0$) at the initial step. Additionally, let us consider that the voltage across the network is constant and equal to $V_0$ and the voltage is connected at time $t=0$. Therefore, initially, the current that passes through $M_1$ is twice the current that passes through $M_2$ ($I_1(0)=2*I_2(0)$). If we wish to compare this step with the initial step of the ant colony algorithm, we may consider that the ants traversing through path 2 get to B slower than the ants that traverse through path 1. Or in other words, the *density* of the ants would be smaller in path 2 compared to path 1.

Getting back to the memristive network, as explained earlier, the current that passes through $M_1$ is twice that of $M_2$. On the other hand, since the rate of change in the memristive devices depends on the current that passes through them, the conductance of $M_2$ changes more quickly compared to $M_1$. Therefore, as time passes, the difference between the conductance of $M_1$ and $M_2$ becomes more pronounced. This increased change in the conductance, results in increase in the difference of the current that passes the two branches. Furthermore, the change in the current resembles the change in the number of ants that traverse path 1 due to increased pheromone after a certain period of time.

The similarity between ant colony algorithm and memristive networks was exploited in [4] to find the solution to a maze. Specifically, the authors explain that the memristive devices should be initialized with a certain resistance and propose connecting the memristive devices using MOSFETs depending on the connections in the maze; however, they fail to explain how the memristive devices should be initialized. Furthermore, they do not consider realistic models based on experimental memristive devices in literature. Besides, they do not consider real models for the MOSFET devices and consider them as ideal switches.

In this paper, we propose using the similarities between memristive networks and ant colony algorithm for image edge detection. To this end, we make the following key contributions:

- We propose a new hardware-friendly algorithm that uses ant colony to perform image edge detection.
- We explain how ant colony algorithm for edge detection can be mapped to a network of memristive devices. For this purpose, we compare different parameters in the ant colony algorithm and explain how they can be represented as physical entities such as voltage, current and memristance of the devices.
- We simulate a memristive network based on the proposed algorithm using MOSFETs in 32nm technology [22] and memristive devices proposed in literature [19] and analyze different design trade-offs regarding energy consumption and performance. Furthermore, we compare our results with the state of the art stochastic circuits implementation of image edge detection. Our results show 28% improvement in the energy compared to a low power CMOS implementation and occupies 5x less area.

The rest of the paper is organized as follows. In Section 2, we propose a hardware-friendly algorithm for edge detection. To this end, we explain how different parameters in the algorithm impact the effectiveness of the algorithm. In Section 3 we propose using memristive devices for implementing the algorithm and explain the impact of various parameters on the hardware complexity of the algorithm. In Section 4, we describe the simulation framework for the proposed memristive implementation. In Section 5, we analyze the simulation results of an implementation of the algorithm using state of the art memristive devices. Finally, Section 6 concludes the paper.

## II. IMPLEMENTATION FRIENDLY ANT COLONY ALGORITHM FOR IMAGE EDGE DETECTION

Ant colony algorithm is based on the search of multiple ants modeled as agents exploring a graph to find the optimum solution to a problem. The graph has nodes (or vertices) and

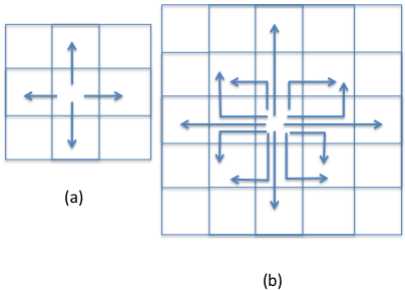

Fig. 2. (a) Path set illustration for L=1.
(b) Path set illustration for L=2.

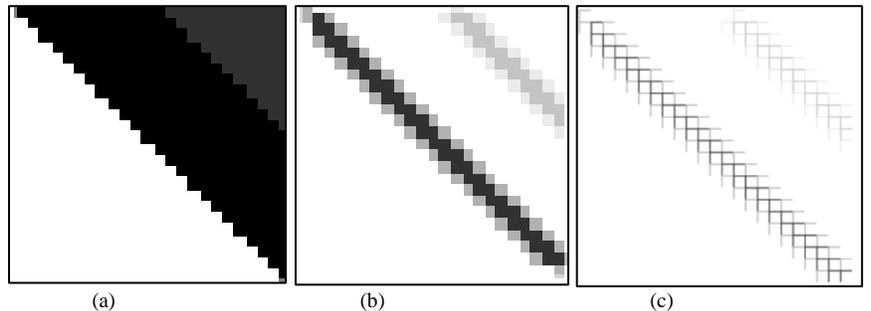

Fig. 3. (a) A sample image with a two edges. (b) The bitmap image of the contrast of the image in (a). (c) The graph representation of the image in (a).

```
Initialize the edges on each pixel
For iteration=1:N
    For i=1:NumColumn
        For j=1:NumRow
            For step=1:L
                Select and go to the next pixel
                Update pheromone
            End //step
        End //j
    End //i
End //iteration
```

Fig. 4. Pseudo-code for the proposed ant colony algorithm.

edges represented as $G=(V,A)$ in which $V$ represents the vertices and $A$ the edges. Each edge connecting nodes $i$ and $j$ has two values associated with it: A heuristic, which defines the favorability of the edge ($d_{ij}$) and a pheromone, which mimics the pheromone in the ant colony system ($\tau_{ij}$).

The ant colony algorithm has four main stages, namely, graph representation, initialization stage, node transition rule and pheromone updating rule.

In order to perform edge detection using ant colony algorithm, there is a need to construct a graph that represents the nature of the problem. In our algorithm, we consider that the image is represented by a two dimensional graph. Furthermore, we consider each pixel as one node and assume that the pixel at $i$th row and $j$th column can be represented as $n_{i,j}$. Furthermore, we consider that there exists an edge between node $n_{i,j}$ and nodes $\{n_{i,j-1}, n_{i,j+1}, n_{i-1,j}, n_{i+1,j}\}$. Therefore, at each node the ant has at most four different choices to make. It also implies that the ants cannot traverse diagonally and can only traverse horizontally and vertically. However, this assumption does not affect the ability to detect diagonal edges because each diagonal edge can be considered as a horizontal step followed by a vertical step. The same assumptions are made in [15] to derive the graph representation.

The next stage is initialization. At this stage, it is required to define the heuristic associated with each edge. For each edge in the graph, the heuristics should define the favorability of the adjacent node. Since the ultimate goal of this algorithm is to detect the edges in the image, the favorability of each node is defined by the contrast of each node. However, the method of defining the contrast varies between different proposed algorithms. In this paper, we define the heuristic associated with each node as:

$$\eta_{(i,j)} = \frac{1}{I_{Max}} \left[ \begin{array}{c} |I(i,j-1) - I(i,j+1)| + \\ |I(i-1,j) - I(i+1,j)| \end{array} \right] \qquad (2)$$

where $I(i,j)$ is the intensity of the pixel at $(i,j)$, $I_{Max}$ is a normalizing factor, set to the maximum intensity variation in the whole image.

The third stage of the algorithm is simulation of ant traversal. Ant traversal is the most complex and time-consuming stage in the algorithm. Therefore, defining an effective, yet implementation-friendly algorithm is of great importance.

We suggest that ants start from each and every pixel in the image. Furthermore, the number of pixels that each ant may traverse is equal to $L$. At the next step, we define the set of all possible "paths" that an ant can traverse as the "path set". Each possible "path" from the initial point of $n_{i0,j0}$ consists of viable sequence of nodes that the ant may traverse without visiting one node more than once. Furthermore, the ant can only traverse to adjacent nodes from each and every node. Furthermore, the number of nodes in each path is equal to $L+1$. For example, if $L=1$, there are 4 paths in the path set. Each of the paths are represented with an index that shows their position in the path set. For example, the paths can be represented as: $\{path_1=\{n_{i0,j0}, n_{i0,j0+1}\}$, $path_2=\{n_{i0,j0}, n_{i0,j0-1}\}$, $path_3=\{n_{i0,j0}, n_{i0+1,j0}\}$, $path_4=\{n_{i0,j0}, n_{i0-1,j0}\}\}$ as shown in Fig. 2(a). As another example, if $L=2$, the paths can be represented as:

$\{ path_1=\{n_{i0,j0}, n_{i0,j0+1}, n_{i0-1,j0+1}\}$, $path_2=\{n_{i0,j0}, n_{i0,j0+1}, n_{i0+1,j0+1}\}$, $path_3=\{n_{i0,j0}, n_{i0,j0-1}, n_{i0-1,j0-1}\}$, $path_4=\{n_{i0,j0}, n_{i0,j0-1}, n_{i0+1,j0-1}\}$, $path_5=\{n_{i0,j0}, n_{i0-1,j0}, n_{i0-1,j0+1}\}$, $path_6=\{n_{i0,j0}, n_{i0-1,j0}, n_{i0-1,j0-1}\}$, $path_7=\{n_{i0,j0}, n_{i0-1,j0}, n_{i0-1,j0-1}\}$, $path_8=\{n_{i0,j0}, n_{i0-1,j0}, n_{i0-1,j0+1}\}$, $path_9=\{n_{i0,j0}, n_{i0-1,j0+1}, n_{i0-2,j0}\}$, $path_{10}=\{n_{i0,j0}, n_{i0+1,j0}, n_{i0+2,j0}\}$, $path_{11}=\{n_{i0,j0}, n_{i0,j0+1}, n_{i0,j0+2}\}$, $path_{12}=\{n_{i0,j0}, n_{i0,j0-1}, n_{i0,j0-2}\}\}$ as shown in Fig. 2 (b).

The next step is to describe the pheromone update rules. To this end, each edge in the graph is considered to have an initial pheromone value ($\tau_{i,j}$). Furthermore, each ant starting at each node chooses the path to traverse based on a combination of the heuristics and pheromone associated with each edge. We consider that the probability of traversing $path_m$ is equal to:

$$p_{path_m} = \frac{\Pi_{(i_t,j_t) \in path_m} \tau_{(i_t,j_t)}^{\alpha} (1/Le_{path_m})^{\beta}}{\sum_{f=1}^{M} \Pi_{(i_t,j_t) \in path_f} \tau_{(i_t,j_t)}^{\alpha} (1/Le_{path_f})^{\beta}},$$

$$Le_{path_m} = \sum_{(i_t,j_t) \in path_m} \eta_{(i_t,j_t)}^{-1} \qquad (3)$$

where $\tau_{i,j}$ is the pheromone leading to node $(i,j)$, $\eta_{i,j}$ is the heuristics associated with node $(i,j)$, $Le_{path_m}$ is the length of $path_m$ and $\alpha$ and $\beta$ are two fitting parameters that define the importance of the heuristics vs. the pheromones.

Once the ant has chosen a path to traverse, the pheromone on that path is updated based on the following rule:

$$\tau_{(i,j)}(k+1) = (1-\rho)\tau_{(i,j)} + \frac{vQ}{Le_{path_m}} \qquad (4)$$

where $\tau_{(i,j)}(x)$ is the pheromone at step $x$. $\rho$ is the pheromone forget rate, $Q$ is a fitting parameter and $Le_{path_m}$ is the length of the chosen path. In other words, the new pheromone value depends on the old pheromone value plus a value that depends on the attractiveness of the path the ant has chosen. For example, larger (smaller) values of $\eta_{(i_t,j_t)}$ result in smaller (larger) path length and thus larger (smaller) pheromones.

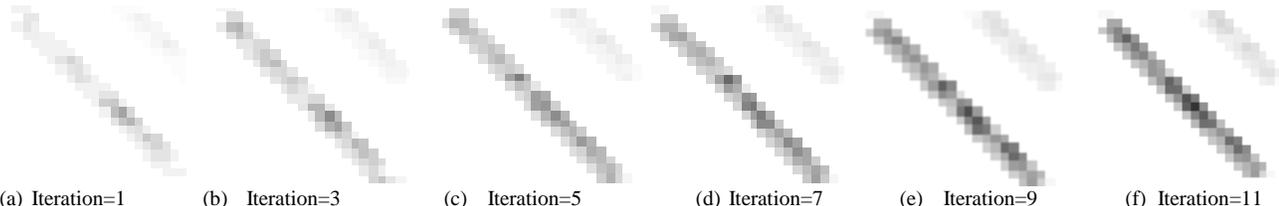

(a) Iteration=1    (b) Iteration=3    (c) Iteration=5    (d) Iteration=7    (e) Iteration=9    (f) Iteration=11

Fig. 5. The amount of pheromone deposited on the pixel map of Fig. 3 (a) as the ant colony algorithm progresses.

In order to illustrate various stages of the algorithm, let us consider the gray-scale image example image shown in Fig. 3 (a). In order to detect the edges of the image, initially, the contrast of each pixel is evaluated and set as the heuristic associated with each pixel as shown in Fig. 3 (b). Specifically, larger (smaller) values of contrast are shown in darker (lighter) gray scale color. The graph representation of the image shown in Fig. 3 (b) is shown in Fig. 3 (c). Observe in Fig. 3 (c) that the edges may have several different values illustrated with different gray-scale color tones. It is noteworthy that this characteristic is different from the maze problem in which the heuristics could possibly have only two distinct values.

Fig. 4 shows the pseudo-code of the ant colony algorithm. There are several important parameters in the algorithm that have to be set correctly. First let us investigate the effect of α and β -- they define the importance of the heuristics vs. the pheromones. For now, we do not wish to emphasize the importance of one over the other. Therefore, the parameter values are set to α=β=1. Furthermore, as we will explain later in Section III, setting these values will ensure an exact correspondence with a memristive implementation.

Another important parameter is the pheromone forget rate ($\rho$). $\rho$ defines how quickly the pheromones evaporate on each path. Setting $\rho$ to higher values results in higher forget rates and results in slower convergence; therefore, $\rho$ is usually set to a small value. Here we set it as $\rho$=0.001. Finally, the ant traversal length (L) should be defined. For the example problem shown in Fig. 3, we have set L=4. We will later elaborate more on L. A code was written in MATLAB to implement the algorithm described as a pseudo-code in Fig. 4. Fig. 5 shows the amount of pheromone deposited on each node as the algorithm progresses. As observed, the pheromones on the edges increase over time compared to pixels without any edge, and the algorithm successfully detects the edges in the image. Although the example in Fig. 3 is an extreme case of edge detection in a gray-scale image with three tones of color, for practical images, the same principles hold.

Now let us get back to analyzing the impact of the ant traversal length on the effectiveness of the algorithm and its complexity. As it can be inferred from Fig. 2, the size of the path set depends on the length of ant traversal. To this end, let us investigate the complexity of the algorithm with respect to the ant traversal length. Furthermore, let us consider that the ant starts its traversal from a node sufficiently far from the image borders. At the first step, the ant can make 4 different choices (up, down, left and right). At the next step, it can make 3 choices (because it cannot go back). At the third step, it can make the same 3 choices; however, the ant cannot traverse in a loop. Therefore, in some cases, it can make only 1 or 2 choices. For example, it cannot make 3 consecutive right turns because it results in a traversal containing a loop. Thus, an upper bound on the number of total choices the ant can make is $4*3^{L-1}$ for a length of L. Note, the complexity of the algorithm increases exponentially with the length of the ant traversal. Hence, from implementation point of view, reducing $L$ is desirable.

Now let us investigate the impact of the ant traversal length on the quality of the detected edges. In order to analyze the effectiveness of our algorithm, test images from USC SIPI database [17] were used as sample images for the implementation. Fig. 6 shows the results of the edge detection with different $L$ for the "Lena" image. As observed, for smaller values of L, the number of edges detected is higher compared to larger values of L. However, regions with high contrast are also represented as edges. These regions are observed as small black dots on the image. On the other hand, for higher values of $L$, the algorithm looks for longer edges. Therefore, very short edges are not detected in the algorithm. Thus, there is a trade-off between noise reduction and detection of short edges. On the other hand, there is a trade-off between the complexity of the implementation and the noise reduction capability. This trade-off raises the question of whether it is possible to benefit from noise reduction in longer ant traversal lengths without significant increase in the implementation complexity. One

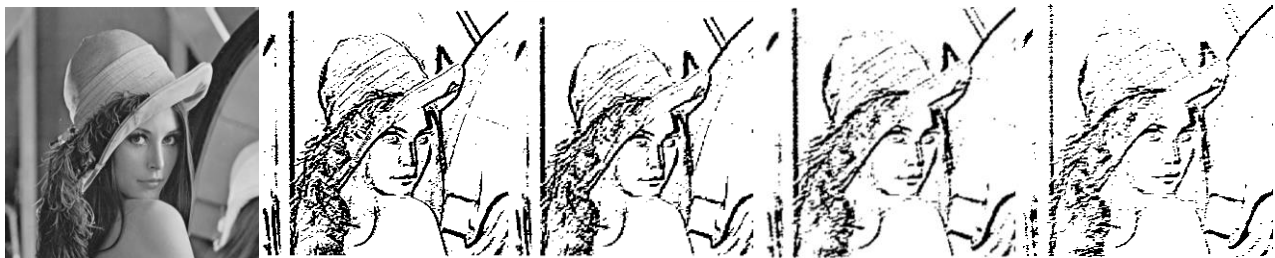

(a) Original image      (b) L=2      (c) L=4      (d) L=6      (e) L=8

Fig. 6. Comparison of the quality of the edges detected for the Lena picture shown in (a) for various lengths of ant traversal (L).

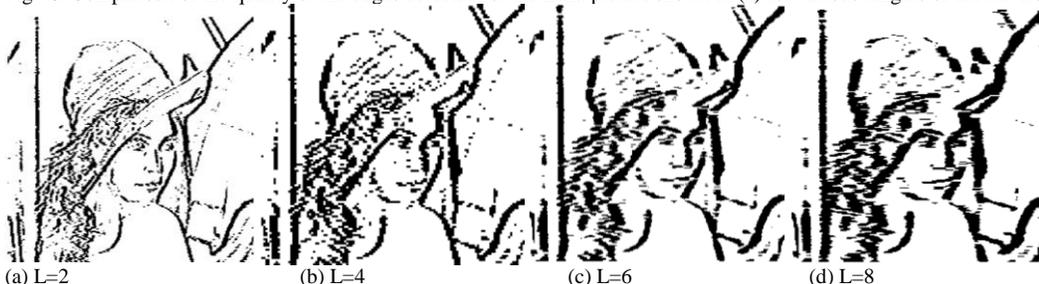

(a) L=2      (b) L=4      (c) L=6      (d) L=8

Fig. 7. Comparison of quality of the edges detected for Lena picture shown in Figure 6 (a) for various lengths of ant traversal with horizontal and vertical patterns only.

viable solution to this problem is to consider only part of the entire path set for large $L$.

For example, if $L=2$, instead of having all 12 paths, we would implement 6 of them and not the others. In other words, the ant could choose only some of the paths and not the others. To this end, we considered implementing only the horizontal and vertical paths and not the others. Fig. 7 shows the edges detected using horizontal and vertical only paths for different lengths of ant traversal. As observed, for smaller values of $L$ the algorithm performs well. However, setting $L> 4$, has a blurring effect on the detected edges. This observation can be explained by considering the fact that at each pixel, the ant may traverse only straight towards one of the four directions around it. Furthermore, it lays pheromone on all of these edges. Setting the ant traversal length too long causes pheromone updates on pixels that are substantially far from the initial pixel of the ant; which causes a blurring effect. Therefore, this solution is only practical in ant traversal lengths that are sufficiently small.

## III. MEMRISTIVE IMPLEMENTATION OF THE ANT COLONY ALGORITHM FOR EDGE DETECTION

In this Section we propose a memristive implementation of the ant colony algorithm. To this end, we will explain how the similarities between memristive devices and the location-based communication of ants can be exploited to implement the algorithm efficiently. To this end, we first investigate a small simple edge detection problem and show how this simple problem can be mapped to a memristive implementation. At the next step, we propose a systematic approach to use the similarities for image edge detection using memristive devices.

### A. A simple edge detection example

In order to show the effectiveness of using ant colony algorithm for image edge detection, let us focus on the progress of the algorithm using a simple example. For this purpose, let us consider the algorithm proposed in Section II for a very small image. Fig. 8 (a-d) illustrate the original and a noisy image sample and their contrast images. If we wish to use the ant colony algorithm for image edge detection in the noisy image in Fig. 8 (c), the first step is to derive the contrast of each pixel as shown in Fig. 8 (d). The next step is to simulate the ant traversal. Let us consider only one ant starting from the center of the image as shown in Fig. 8 (e). Also, let us set the ant traversal length to $L=4$ and the pheromones on all the pixels $\tau_0$. Also, let us consider that there are only purely horizontal and purely vertical paths in the path set. In other words, the ant can traverse to up, down, left or right directions for 4 steps. Furthermore, let us consider that the pixels with a white color have $\eta_0=1$, the ones in grey have $\eta_1=5$, $\eta_2=10$ and $\eta_3=15$ depending on their intensity as illustrated in Fig. 8 (e). Under such conditions, the length associated with each path can be written as:

$$Le_{up} = Le_{left} = \frac{4}{\eta_0} = 4, Le_{right} = \frac{1}{\eta_0} + \frac{3}{\eta_2} = 1.3,$$
$$Le_{down} = \frac{2}{\eta_0} + \frac{1}{\eta_1} + \frac{1}{\eta_3} = 2.266 \quad (5)$$

Observe in Eq. 5 that the length of ant traversal to up and left directions is substantially larger than the length of the ant traversal to right and down directions. In order to simplify the example, let us consider the length of ant traversal to up and left directions to be infinity and the probability of traversal to these two directions to be zero. On the other hand, the probability of traversing to the right and down directions can be written as:

$$p_{right} = \frac{(\tau_0^4)^\alpha (1/Le_{right})^\beta}{(\tau_0^4)^\alpha (1/Le_{right})^\beta + (\tau_0^4)^\alpha (1/Le_{down})^\beta},$$
$$p_{down} = \frac{(\tau_0^4)^\alpha (1/Le_{down})^\beta}{(\tau_0^4)^\alpha (1/Le_{right})^\beta + (\tau_0^4)^\alpha (1/Le_{down})^\beta}, \quad (6)$$

Where $\tau_0$ is the initial pheromone on each node, $Le_d$ is the length of ant traversal to direction $d$ and $\alpha$ and $\beta$ are two fitting parameters. Now let us consider the path to the right as *path 1* and the path downward as *path 2*. Additionally, the pheromones of the paths can be represented as the product of the pheromones on all of the constituent nodes in each path. Therefore, at the first time step, we have:

$$\tau_1 = \tau_2 = \tau_0^4 \quad (7)$$

Rewriting Eq. 3 considering Eq. 6, 7 results in:

$$p_{1(2)} = \frac{\tau_{1(2)}^\alpha (1/Le_{1(2)})^\beta}{\tau_1^\alpha (1/Le_1)^\beta + \tau_2^\alpha (1/Le_2)^\beta} \quad (8)$$

in which $\tau_{1(2)}$ is pheromones laid on each path and $Le_i$ is the length of ant traversal in the $i$th path. Besides, the pheromone dynamics on the first (second) path is:

$$\tau_{1(2)}(k + 1) = (1 - \rho)\tau_{1(2)}(k) + \nu Q / Le_{1(2)} \quad (9)$$

Where $\rho$ is the pheromone forget rate and $\nu$ and Q are two fitting parameters. Now let us assume that the number of ants entering the image has a constant rate, $\gamma$. Then, the amount of ants added within a time interval $dt$ is equal to $\gamma dt$. Therefore, Eq. 9 can be rewritten as [4,5]:

$$\frac{d\tau_{1(2)}}{dt} = -\gamma\rho\tau_{1(2)} + p_{1(2)}\gamma\nu Q / Le_{1(2)} = \quad (10)$$
$$-\gamma\rho\tau_{1(2)} + \frac{\gamma\nu Q}{Le_{1(2)}} \frac{\tau_{1(2)}^\alpha (1/Le_{1(2)})^\beta}{\tau_1^\alpha (1/Le_1)^\beta + \tau_2^\alpha (1/Le_2)^\beta}$$

In order to implement the ant colony algorithm using memristive devices, let us consider that a memristive device is used to represent each path in the path set as shown in Fig. 8 (e). Also, let us consider that the conductance of each memristive device can be represented as:

$$G_d(x) = G_{on\_d} * x + G_{off\_d} * (1 - x), G_{on\_d} > G_{off\_d} \quad (11)$$

where $G_d(x)$ is the conductance of the memristive device, $G_{on\_d}$ is the conductance of the memristive device in the ON state and $G_{off\_d}$ is the conductance of the memristive device in the OFF state and $x$ is the internal variable of the memristive device. Besides, let us consider that the equation for the internal variable should contain a drift term ($K$) that formulates the dependence of the internal state on the current passing through it as well as a relaxation term ($\xi$):

$$\frac{dx}{dt} = KI(t) - \xi x \quad (12)$$

in which $K$ is the drift constant and $\xi$ is the relaxation term.

Also, the initial conductance of all of the memristors is equal to $G_{off\_d}$ where $d$ can take four values: up, down, left and right. Besides, the value of $G_{off\_d}$ is inversely proportional to the length of each path:

$$G_{off\_up} = G_{off\_left} = 1/4, G_{off\_right} = 1/1.3,$$
$$G_{off\_down} = 1/2.266 \quad (13)$$

In order to analyze the similarities between the memristive network and the ant colony algorithm, there is a need to analyze the dynamics of the circuit shown in Fig. 8 (f) with that of the ant colony algorithm. Observe in Eq. 13 that $G_{off\_up}$ and $G_{off\_left}$ are very small. In order to simplify the illustration and keep the correspondence with the ant colony algorithm example, let us consider that $G_{off\_up}=G_{off\_left}=0$. Therefore, they are considered to be open circuit. The circuit implementation of the right and down direction paths in Fig. 8 (e) are illustrated in Fig. 8 (f). Furthermore, let us name $G_{right}$ as $G_1$ and $G_{down}$ as $G_2$. The current passing through each branch illustrated in Fig. 8 (f) can be written as:

$$I_{1(2)} = I_0 \frac{G_{1(2)}}{G_1 + G_2} \qquad (14)$$

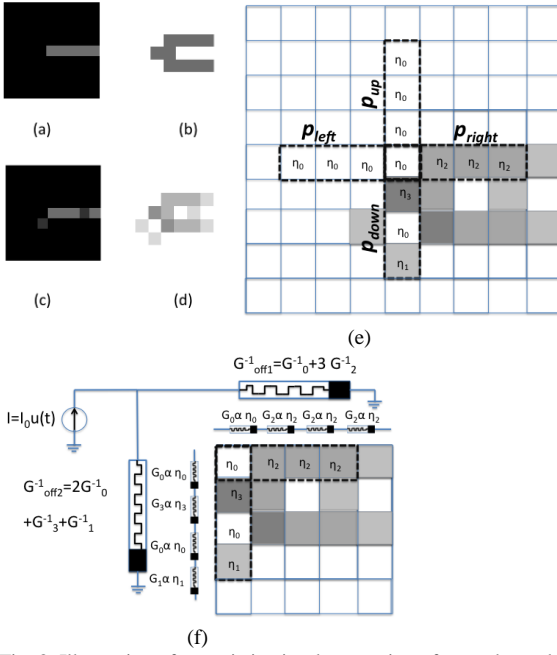

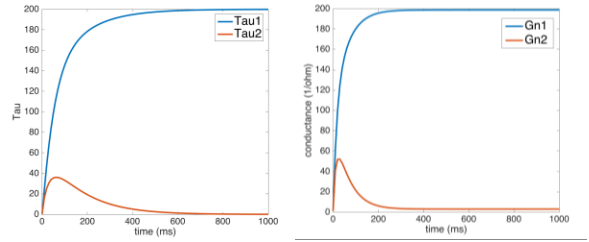

Fig. 9. Comparison of pheromone values and the normalized conductance in Eq. 6 and 15 respectively. (a) parameters used are $\gamma=20$, $\rho=1$, $L_1=1.3$, $L_2=2.266$, $\tau_1(0)=\tau_2(0)=0.01$, the parameters are adjusted for illustration purposes. (b) $I_0=1$, $\xi=50$, $G_{off1}=2.266$, $G_{off2}=1.3$, $G_{on1}=2200$, $G_{on2}=1300$, $Gn_1(0)=Gn_2(0)=1$ the parameters are adjusted for illustration purposes only.

Fig. 8. Illustration of memristive implementation of ant colony algorithm for a small image. (a,b) original image and its edges. (c,d) noisy image and its edges. (e) probability of ant traversal for all of the paths in the path set. (f) Memristive implementation of (e).

On the other hand, in order to analyze the dynamics of the network, we assume $G_{off1}=G_{off\_right}$ and $G_{off2}=G_{off\_down}$ as initial conditions. Eventually, rewriting Eq. 12 considering Eq. 14, the normalized conductance of each branch and considering $Gn_i(t)=G_i/G_{offi}$ can be written as:

$$\frac{dGn_{1(2)}}{dt} = -\xi\left(Gn_{1(2)}-1\right)$$
$$+ KI_0 \left(\frac{G_{on1(2)}}{G_{off1(2)}}-1\right) \frac{Gn_{1(2)}(t)G_{off1(2)}}{Gn_1(t)G_{off1}+Gn_2(t)G_{off2}} \qquad (15)$$

Comparing Eq. 15 and 6, it can be concluded that the ant colony algorithm is implemented in Eq. 15 with parameters $\alpha=1$ and $\beta=1$. Furthermore, the initial off ($G_{off1(2)}$) state can be interpreted as the heuristics associated with each path ($\eta_{1(2)}$). Besides, the normalized conductance $Gn_i(t)=G_i/G_{offi}$, which is proportional to the internal state variable, plays the role of the pheromone strength $\tau_{1(2)}$. Fig. 9 (a) and (b) illustrate the value of pheromones in Eq. 6 and the value of the normalized conductance in Eq. 15 respectively. Observe in Fig. 9 that the pheromones and the normalized conductance show a similar behavior and settle to a final state similarly. Furthermore, observe in Fig. 9 (a) that although there is a high contrast pixel in the downward direction, the pheromones on this path settle to a small state showing that this path is not an edge. Therefore, the algorithm successfully distinguishes between a real edge and that caused by a noisy pixel. Similar assumptions can be made for the memristive implementation in Fig. 9 (b).

However, there are differences between these two systems such as the final relaxation state. In the ant colony system, the final state is zero for undesired paths; however, in the memristive implementation, the final relaxation state is a small positive non-zero number. Despite, all these differences, it has been shown that these two systems come to the same solution [4,9-12].

On the other hand, it has been shown that similar results are achievable if the current source would be replaced by a voltage source. For example, if a voltage is applied to a memristive network representing a maze, the final solution can be obtained in a similar fashion [9-12].

Additionally, the aforementioned proof is only viable for one ant traversing from a specific pixel; nevertheless, the traversal of several ants starting from various locations in different orders is not analyzed. Finally, the proof provided in Eq. 11-15 can be rewritten for voltage-based memristive devices and similar results can be obtained using these devices. Specifically, using source conversion, all of the series connections should be changed to parallel ones and the current source should be transformed to voltage source. In the next subsections, we will propose a systematic method for image edge detection using voltage based memristive devices based on ant colony algorithm.

### B. Graph mapping to memristive network

Every ant colony problem is represented as a graph explaining the nature of the problem. In order to solve the problem using memristive devices, the graph should be mapped to a memristive network. To this end, we consider that each and every pixel is represented as a memristive device. Furthermore, we assume that the memristive device at $i$th row and $j$th column can be represented as $Me_{i,j}$. At the next step, we consider that $Me_{i,j}$ may be connected to $\{Me_{i,j-1}, Me_{i,j+1}, Me_{i,j-1}, Me_{i-1,j}, Me_{i+1,j}\}$. Fig. 10 (a) illustrates the required circuitry for each and every pixel. The circuitry consists of the memristive element ($Me_{i,j}$), initialization circuitry, which is used to initialize the memristive device, ant traversal simulation circuitry which is used to simulate the ant traversal and the read circuitry, which is used to read out the value of the memristive device once the stopping criterion is reached. In the following Subsections, we will explain each circuitry with respect to its functionality.

### C. Initialization of the memristive network

The main goal of the initialization step is to program the memristive devices based on the definition of the problem. As explained earlier in Eq. 13, the initial conductance of the device defines the favorability of each pixel with respect to the edge detection problem. Therefore, the initial value of the conductance is proportional to the contrast of each pixel as explained in Eq. 2:

$$G_{ini(i,j)} \propto \eta_{(i,j)}^{-1} \qquad (16)$$

where $G_{ini(i,j)}$ is the initial conductance of the memristor.

Activating the initialization circuitry for each pixel performs the initialization step. For this purpose, $M_{ini}$ is used to connect the memristive device to the initialization circuitry. Furthermore, the source-line ($SL$) is pulled up to $V_{dd}$. On the other hand, the amount of time $M_{ini}$ is ON defines the value of $G_{ini}$. Specifically, longer (smaller) ON times result in larger (smaller) changes in the internal variable. Therefore, the ON time should be adjusted according to the value of each pixel.

### D. Ant traversal

At the next step, ant traversal is simulated. Ant traversal includes mapping the traversal rules and pheromone update rules to the connections and sequence of operation in the memristive network.

In order to simulate node transition, we consider connecting proper memristive devices to other memristive devices and to the power source(s). Specifically, we assume that the length of the traversal for each ant is $L$. As explained in Section II, the ant may traverse through different *paths* in the

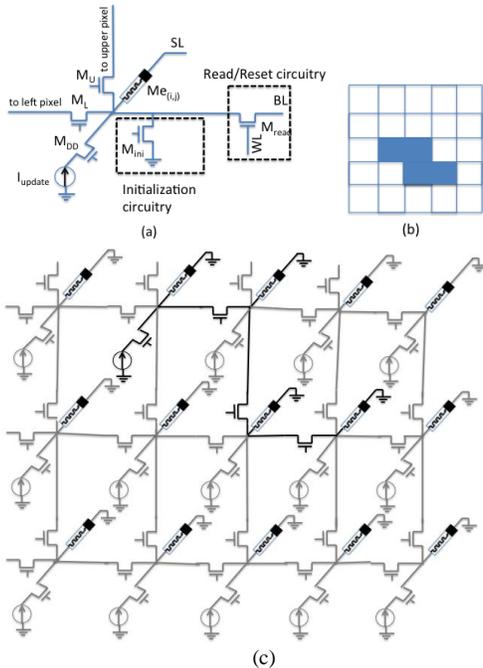

*path set.* In order to simulate ant traversal through each *path,* memristors are connected in one of the *paths*.

In order to simulate pheromone update; it is considered that each ant traverses a specific path at a time. As explained in Eq. 15, the change in the value of the internal variable in the memristive device is interpreted as the change in the pheromone value. Therefore, the memristive devices at each path are connected to a current source.

The current source causes a change in the internal variable of the memristive element. Ant traversal circuitry is used to realize the connections. Observe in Fig. 10 (a) that the ant traversal simulation circuitry consists of three transistors. Notably, $M_L$ and $M_U$ are used to connect each memristive device to the adjacent memristive devices horizontally and vertically. Furthermore, $M_{DD}$ is used to connect the devices to the current source $I_{update}$. This current source is used to change the internal state of the memristive device.

Fig. 10 (c) shows the connections required for the pattern shown in Fig. 10 (b). Specifically, the wires shown in black are conducting and the ones in grey are disconnected. Observe in Fig. 10 that the source-line ($SL$) is grounded during the ant traversal simulation.

The ant traversal is simulated for a single path at a time. Furthermore, it is considered that ants start traversing the image in non-overlapping patterns. The reason for this consideration is that the ant traversal can be simulated in a massively parallel fashion throughout the image. For example, let us consider that the ant traversal length is equal to three pixels and we wish to simulate a horizontal pattern of three pixels. The ant traversal for all of the ants in the image is performed in three phases. In the first phase, we consider that ants start their traversal from pixels at $\{(i,9j+1),(i,9j+4),(i,9j+7)\}$ columns only and they traverse to the right. Therefore, memristive devices are connected in three $\{\{(i,9j+1), (i,9j+2), (i,9j+3)\}, \{(i,9j+4), (i,9j+5), (i,9j+6)\}, \{(i,9j+7), (i,9j+8), (i,9(j+1))\}\}$ where $i$ is the row of each pixel and $9j+x$ is the column of the pixel. In the second phase, it is considered that the ants start from the pixels at $\{(i,9j+2),(i,9j+5),(i,9j+8)\}$ and traverse to the right: They are connected in groups $\{\{(i,9j+2), (i,9j+3), (i,9j+4)\}, \{(i,9j+5), (i,9j+6), (i,9j+7)\}, \{(i,9j+8), (i,9(j+1)), (i,9(j+1)+1)\}\}$. In the third phase, it is considered that the ants start from the pixels at $\{(i,9j+3),(i,9j+6),(i,9(j+1))\}$ and traverse to the right. Therefore, they are connected in groups $\{\{(i,9j+3), (i,9j+4), (i,9j+5)\}, \{(i,9j+6), (i,9j+7), (i,9j+8)\}, \{(i,9(j+1)), (i,9(j+1)+1), (i,9(j+1)+2)\}\}$. Fig. 11 shows the ant traversal simulation for a purely horizontal pattern of three pixels for the two different phases.

In order to sweep all of the design space, different paths are simulated consecutively for the entire image. Furthermore, we should emphasize that although we consider the same path for each and every pixel, the amount of change in the internal variable of the device depends on the value of each memristive device. Furthermore, since this value mimics the pheromone deposit, the amount of pheromone laid on each pixel is different and depends on the location of the pixel.

### E. Stopping criterion, read-out and reset

The stopping criterion is reached once a certain number of ant traversals have been performed. The number of traversal updates is defined by trial and error and the desired quality of the detected edges.

Once the stopping criterion is reached, the conductance of each memristor representing each pixel should be sensed.

Fig. 10. (a) Implementation of each pixel for voltage based memristive devices. (b) A sample pattern of ant traversal. (c) Illustration of connections of adjacent pixels for voltage based memristive devices.

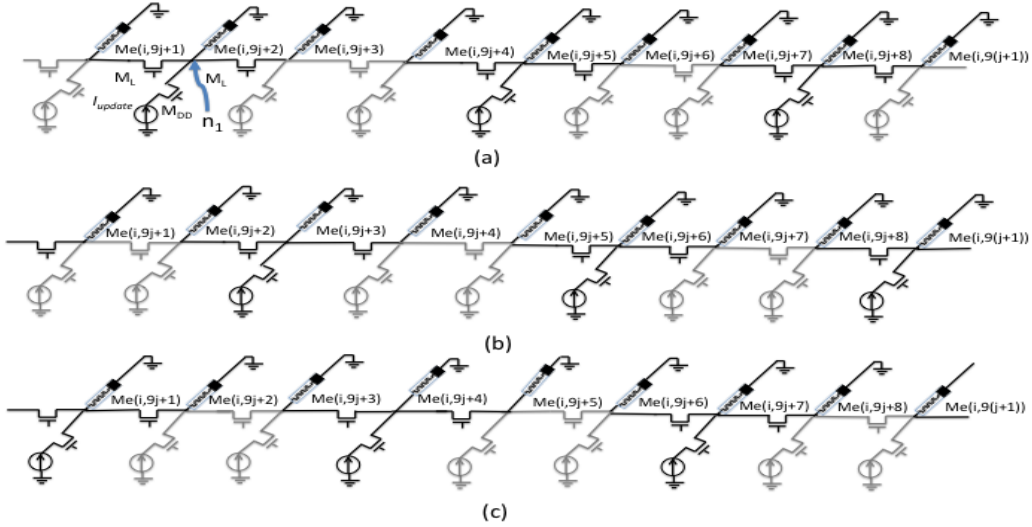

Fig. 11. Illustration of ant traversal simulation for a purely horizontal patter of length L=3. (a) Illustration of the first ant traversal phase. (b) Illustration of the second ant traversal phase. (c) Illustration of the third ant traversal phase.

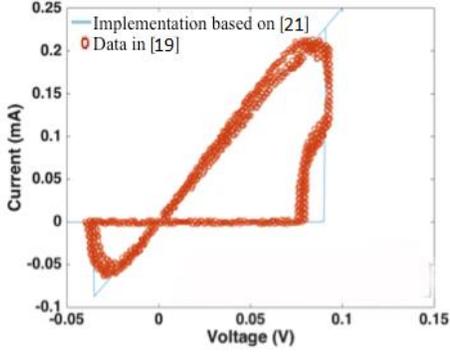

Fig. 12. Comparison of the implementation of model in [21] with the experimental data in [19].

Activating the read circuitry performs the sensing of the resistance and the final read-out. For this purpose, the word-line (*WL*) on each line is activated and the bit-line (*BL*) is pre-charged to a small voltage and the *SL* is grounded. Eventually, the *BL* is sensed using a current sense amplifier and the edges are derived.

Finally, once the values of the memristive devices are read out, there is a need to reset all of the devices to ensure correct analysis of consecutive images. In order to reset the devices, a voltage is applied to the BL, SL is grounded and the WL is activated. The voltage causes the memristive device to be reset to its original OFF state.

## IV. SIMULATION FRAMEWORK FOR MEMRISTIVE IMPLEMENTATION

A simulation framework was developed to investigate edge detection using memristive networks based on swarm intelligence. The simulation framework consists of four main modules: the memristive device simulation module, the initialization simulation module, the ant traversal simulation module and the read-out/reset module.

### A. Memristive device simulation module

At first, the memristive device simulation module was developed. In order to have a realistic analysis of the algorithm, there was a need to choose a memristive device. There are several memristive devices proposed in literature, e.g. [18,19]. Each of these devices has various characteristics that make them suitable for different applications.

There are several different issues that play important roles in defining the devices suitable for our application. The first important factor is the ability to integrate with CMOS.

The second important factor is the conductance of the memristive device. Observe in Fig. 11 that the MOS transistors are used as switches to power gate the memristive devices. Furthermore, as explained earlier, the effectiveness of the algorithm depends on the change in the voltage when the conductance of the memristors changes. Therefore, the $r_{ds}$ of the NMOS should be sufficiently smaller than that of memristive devices to ensure correct operation of the algorithm.

The third important factor is the difference between the ON conductance and the OFF conductance of the device. Since we are considering continuous and gradual change in the conductance and the current passing through the network to change in accordance with the conductance, higher difference between the ON and the OFF state is desired.

The fourth important factor is the drift constant. The drift constant, defines the rate at which the internal variable changes with respect to the applied voltage. The drift constant plays an important role in the performance and the energy consumption of the network. Larger drift constant results in faster change in the internal variable for a fixed voltage across the device. Therefore, it contributes to the speed of operation. On the other hand, the energy consumption depends on the applied voltage and the time required for each update.

The fifth factor is the relaxation factor. The relaxation factor defines the rate at which the memristive device looses its value, which in turn, corresponds to the evaporation rate of pheromones. In general, the desired relaxation factor depends on the algorithm. Note that not all memristive devices in literature have relaxation factors and this fact should be considered during the design.

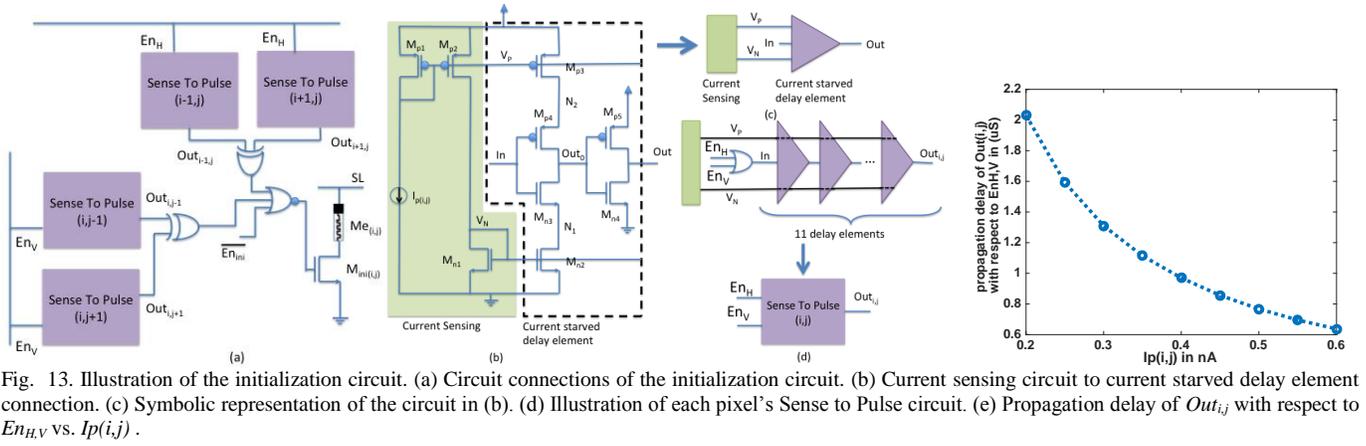

Fig. 13. Illustration of the initialization circuit. (a) Circuit connections of the initialization circuit. (b) Current sensing circuit to current starved delay element connection. (c) Symbolic representation of the circuit in (b). (d) Illustration of each pixel's Sense to Pulse circuit. (e) Propagation delay of $Out_{i,j}$ with respect to $En_{H/V}$ vs. $Ip(i,j)$ .

TABLE 1. PARAMETERS USED TO OBTAIN FIG. 12 BASED ON OUR IMPLEMENTATION OF THE MODEL IN [21].

| Parameter | Value |
|---|---|
| Roff | 1 MΩ |
| Ron | 400 Ω |
| Vtp | 80e-3 |
| Vtn | -35e-3 |
| $\beta_p$ | 19.6e3 |
| $\beta_n$ | 17.5e3 |

TABLE 2. SIMULATION RESULTS OF THE INITIALIZATION CIRCUITRY MODULE

| Parameter | Value |
|---|---|
| Area | 24.356 μm² |
| $V_{dd}$ | 1.05 V |
| Power | 1 μW ~ 22 μW |
| Duration of each programming pulse ($En_H, En_V$) | 2 μS |
| Number of pulses each direction | 2 |
| Energy consumption for each programming pulse | 6 pJ ~ 132 pJ |
| $I_{p(i,j)}$ | 50 pA ~ 1nA |

The sixth factor, is the type (current based or voltage based) of the memristive device. Due to the perceptible similarities between current based memristive devices and the traversal of ants, they have been used to implement swarm based memristive networks. However, voltage based memristive devices can also be used to implement memristive networks using ant colony as explained in Section 3. To this end, ideally, the source transformation of the circuits can be used for realization of the memristive network from one type of device to the other. As an example, current based memristive devices should be connected in series to mimic a path while voltage based memristive elements should be connected in parallel to mimic a path.

Although theoretically, either of these two types of memristive devices is not preferred over the other, practically, voltage based memristive devices are favorable over current based ones. The main reason for this is the parallel connection of these devices to mimic the ant traversal. The parallel connection prevents stacking of several MOS transistors, used as switches, to ensure their proper operation.

Considering different factors mentioned above, the device in [19] was considered in our work. Furthermore, it was modeled in accordance with the model explained in [21] with different threshold voltages for the positive and negative voltages:

$$f_m = \begin{cases} \beta_p(V_m - V_{tp}) & V_m > V_{tp} \\ -\beta_n(V_m - V_{tn}) & V_m < V_{tn} \\ 0 & V_{tn} < V_m < V_{tp} \end{cases} \quad (18)$$

$$\frac{dx}{dt} = f_m, \quad R = R_{off}(1-x) + R_{on}x$$

where $V_m$ is the voltage across the memristor, $V_{tp}$ ($V_{tn}$) are the positive (negative) threshold voltages and $\beta_p$ ($\beta_n$) are the drift constant for positive (negative) voltages. Also, $R_{off}$ is the resistance of the memristors in the off state and $R_{on}$ is its resistance at the on state. Finally, $R$ is the resistance of the device and $x$ is its internal variable. The model was evaluated in MATLAB and the results were compared against the experimental data in [19]. Fig. 12 compares the results obtained by the model in [21] with the data published in [19]. As observed, the simulation results of our model are in close agreement with the experimental data. The parameters used to obtain these results are shown in Table 1.

The memristive device simulation module is used in all the other modules explained in Subsections B, C and D for transient simulation of the circuit. To this end, the entire circuit is considered with respect to the memristor and the differential equation in Eq. 18 is solved self consistently. Specifically, at each time step of the transient simulation, the resistance of the memristor is derived based on the current internal variable and the connections in the circuit. Eventually, the voltage and current of each component in the circuit is derived. Furthermore, the value of the internal variable is updated based on the voltage of the memristor derived at each time step.

### B. Initialization circuitry module

The initialization circuitry is responsible to initialize the memristive devices to the contrast of each pixel based on Eq. 2. Changing the state of the internal variable of the memristive device requires applying a voltage to the device for a certain amount of time. Changing either the voltage or the amount of time the voltage is applied to the memristive device could potentially impact the internal variable of the device. Therefore, the values of the contrast of each pixel could be encoded into the voltage or the amount of time the initialization takes place. Our research and analysis shows that changing the latter is more efficient in terms of energy consumption and performance.

The ant traversal was simulated similar to what was explained in Section III (d) with some modifications. In Section III, we considered the MOS transistors as ideal switches and did not consider the impact of the MOS parameters on the correctness of the implementation. As an example, let us consider the connections in Fig. 11. Observe in Fig. 11 (a) that $Me(i,9j+2)$ is connected to $I_{update}$ through $M_{DD}$ only; however, $Me(i,9j+1)$ and $Me(i,9j+3)$ are connected to $I_{update}$ through the series of two transistors $M_{DD}$ and $M_L$. Therefore, the three memristive devices ($Me(i,9j+1)$, $Me(i,9j+2)$, $Me(i,9j+3)$) are not equal with respect to $I_{update}$. In other words, if the KCL equation is written for node $n_1$, we have:

$$I_{update} = \frac{V_{n_1}}{R_{Me(i,9j+2)}} + \frac{V_{n_1}}{R_{Me(i,9j+1)} + R_{dsM_L}} + \frac{V_{n_1}}{R_{Me(i,9j+3)} + R_{dsM_L}} \quad (17)$$

where $V_{n_1}$ is the voltage at node $n_1$, $R_{Me(x,y)}$ is the resistance of the memristive devices at location $(x,y)$. Also, $R_{dsM_L}$ is the drain source resistance of $M_L$. Observe in Eq. 17 that the effective resistance of the two branches of $(i,(9j+1))$ and $(i,9j+3)$ are different due to the existence of the $M_L$ transistor. This structural mismatch between the two paths causes disproportionate change in the internal variable in the two adjacent pixels. Furthermore, if the number of the memristive devices in the path increases, this mismatch becomes more pronounced.

In order to solve this problem, each distinct path is implemented using unique transistors. This method of connection ensures symmetric connection to all of the memristive devices in the path. Fig. 16 shows a sample connection of memristive devices using the symmetric connection system for the length of $L=3$. Observe in Fig. 16 that if $\phi_{1l}$ is enabled, the first three memristive devices are connected to $I_{update}$. However, if $\phi_{2h}$ is enabled, the second memristor is connected to the third and the fourth. In order to simulate the ant traversal, signal $\phi_{ih}$ is enabled followed by $\phi_{iv}$. The ant traversal simulation is performed in several iterations. Each iteration consists of activating the six different ant traversal signals as illustrated in Fig. 15. Table 3 shows the simulation results for the ant traversal simulation module normalized to each pixel.

### D. Read-out/Reset circuitry module

The read/reset circuitry consists of transistors used to read out the state of the memristive device as well as resetting them to the original state.

In order to read the value of the memristive device, the $WL$ is pulled up to $V_{dd}$ and the $BL$ is pulled up to a small voltage and $SL$ is grounded. At the next step, the current is sensed using a current sense amplifier. Note, passing current through the memristive device could potentially change its internal state. Therefore, $M_{read}$ is designed such that the voltage applied to the device would be less than the threshold voltage of the device.

Note that the read operation should be performed for each row separately.

In order to reset the device, $WL$ is enabled and $BL$ is pulled up to $V_{dd}$ and $SL$ is grounded. The reset is performed for all of the memristive devices simultaneously. At the end of this step, the devices are reset back to the minimum conductance state. Table 4 shows the simulation results for the Read-out/Reset circuitry module.

## V. SIMULATION RESULTS

The simulation framework was used to simulate the dynamics of the memristive network. In general, the energy consumption and the termination condition of the algorithm depend on the image. Herein, we provide the results for a case study of the "pepper" image [17]. At first, each pixel should be initialized as explained in Section 4. Theoretically, the value of the initialization resistance of each pixel should be proportionate to the value of its heuristic; this value is proportionate to the contrast of each pixel as explained in more detail in Section 2. However, the exact value should be defined in the simulation. On the other hand, practical memristive devices have a valid dynamic range in which their characteristics are valid. Furthermore, the change in the internal variable is expensive in terms of performance and power. Therefore, setting it to lower values is desirable. Nevertheless, smaller values of conductance decreases the noise immunity. Our simulation results show that increasing the initialization resistance over roughly 15% higher than the dynamic range between $R_{on}$ and $R_{off}$ does not improve the noise immunity. Therefore, in our simulation framework, we considered a maximum of 15% increase in the resistance value of each pixel (after the initialization step, the resistance of the memristors would be $12.5\text{k}\Omega \sim 150\text{ k}\Omega$). In order to obtain this setting, each pixel was initialized using two initialization pulses in each direction. The pulse duration was set to be 2 µs as explained in Section 4.

Theoretically, the ant traversal should enhance the results obtained by the algorithm or saturate to the final edges detected in the image. However, it is always desired to stop the simulation as soon as the results are obtained. This stopping could potentially reduce the energy consumption and enhance the performance. On the other hand, in a realistic memristive implementation, there is an extra factor that comes into picture: the saturation of the memristors. Observe in Fig. 12 the memristive device could potentially have two bounded values. Once the resistance associated with a pixel has reached the final value it saturates and stays at that position. On the other hand, the adjacent pixels do not saturates and are still affected by the ant traversal simulation. Therefore, the circuitry does not perform the intended task of changing the conductance for some

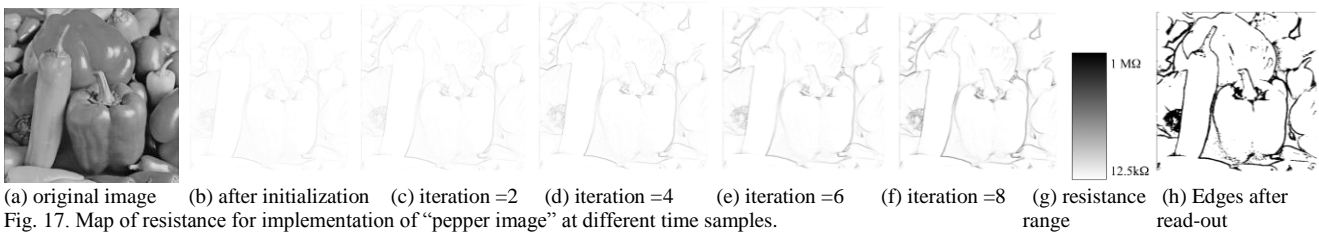

(a) original image    (b) after initialization    (c) iteration =2    (d) iteration =4    (e) iteration =6    (f) iteration =8    (g) resistance range    (h) Edges after read-out

Fig. 17. Map of resistance for implementation of "pepper image" at different time samples.

Table 5. Comparison of image edge detection implementation with CMOS implementations

| Implementation | Area (μm²) | Delay (μS) | Energy (nJ) |
|---|---|---|---|
| SC* in [23] | 4312 | 1.3 | 28.34 |
| SC* in [24] | 200 | 0.058 | 1.14 |
| This work | 37.22 | 68 | 0.819 |

*Stochastic Circuits implementation

of the pixels while it still performs well on the others. This undesired selectivity of the circuitry causes image distortion. Based on our simulation results, the ON time for each ant traversal was considered to be 1 μs. Our simulations show that the total time required to reach the final state is 60 μs. Fig. 17 shows the resistance of the memristors associated with the pixels of the "pepper" image at different simulation times. The number of pixels considered for this implementation is 512x512 pixels. Furthermore, the energy consumption of the implementation is equal to 0.819 nJ per pixel including the reset energy.

We considered our implementation under non-ideal conditions. As explained in Section 3, bio-inspired algorithms have an inherent immunity to noise. For this purpose, we considered the variations in the form of added noise to the input signal and added a uniform noise to the value of the pixels. Fig. 18 shows the image and the detected edges for different percentage of noise. Observe in Fig. 18 that our method generates acceptable results for noise levels up to 30% of the original value.

In order to have a fair comparison with a CMOS implementation, we compared our implementation with some of the state of the art implementations in literature. Recently, it has been shown that image edge detection can be performed using stochastic circuits [23,24] very efficiently. To this end, in [23,24] the authors have simulated custom implementation of these algorithms in hardware. Table 5 compares our results with these implementations. Observe in Table 5 that our implementation consumes less energy compared to the other two implementations. Furthermore, we would like to emphasize that the data reported in [23,24] does not contain the energy required for the digitization process. It is assumed that the data is already digitized. Therefore, the realistic value of energy consumption for the CMOS implementation is larger than what is reported in [23,24]. On the other hand,

However, our implementation is orders of magnitude slower than CMOS. The reason for this poor performance is the slow change in the internal variable of the memristive devices. We believe that with future advancements in the fabrication of memristive devices, the performance of our methodology can be improved.

## VI. Conclusion

In this paper, we proposed usage of memristive networks for image edge detection based on swarm intelligence. To this end, we proposed a hardware implementation friendly ant colony algorithm for image edge detection. At the next step, we proposed an implementation of the algorithm using memristive devices. Finally, we developed a simulation framework to evaluate our proposed implementation strategy. We implemented the algorithm using state-of-the-art memristive devices. Our results show that our implementation consumes about 5X less area compared to a CMOS implementation of edge detection algorithm. Also, our implementation consumes up to 28% less energy; however, it has three orders of magnitude worse performance. We believe that future advancements in the fabrication of memristive devices could potentially improve the performance of our proposed methodology.

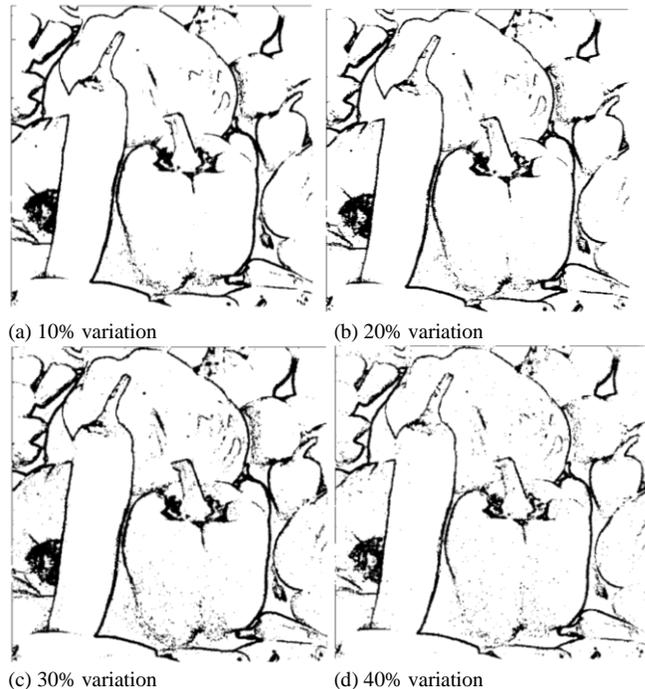

(a) 10% variation  (b) 20% variation

(c) 30% variation  (d) 40% variation

Fig. 18. Illustration of image edge detection using the proposed framework for different levels of variations in the intensity values.